# Integrating AI and Quantum-Inspired Techniques for Efficient Enzyme Fermentation Optimization

Ying-Wei Tseng[1], Yu-Ting Kao[1], Yeong-Jar Chang[1], Chia-Ho Ou[2,3], Wen-Chih Chang[4],
Jin-Jia Wang[5] and Yung-Hsiang Lin[5]

*Electro-Optical Systems Laboratory (EOSL), Industrial Technology Research Institute, Taiwan*[1]
*Department of Computer Science and Information Engineering, National Pingtung University, Taiwan*[2]
*Graduate School of Information Sciences, Tohoku University, Japan*[3]
*International Master Program in Information Technology and Applications, National Pingtung University, Taiwan*[4]
*TCI co, Ltd, Taiwan*[5]

***Abstract***—This paper introduces a new method that combines Artificial Intelligence (AI) and quantum-inspired techniques to improve the efficiency of multi-variable optimization experiments. By using advanced software simulations, this approach significantly reduces the time and cost compared to traditional physical experiments. The research focuses on enzyme fermentation, demonstrating that this method can achieve better results with fewer experiments. The findings highlight the potential of this approach to more effectively identify optimal formulations, leading to advancements in enzyme fermentation and other fields that require complex optimization. Initially, the Active Ingredients (AIN) could not be improved even after 600 experiments. However, by adopting the method outlined in this paper, we were able to identify a better formula in just 405 experiments. This resulted in an increase of AIN from 8481 to 10068, representing an improvement of 18.7%.

## I. INTRODUCTION

In contemporary scientific and industrial contexts, optimizing complex systems through combinatorial approaches poses a significant challenge. This is particularly crucial in biotechnology, where multi-variable experiments, such as enzyme fermentation, are pivotal. Variables like temperature, stirring frequency, pH value, tryptophan, brown rice flour, and nutrient concentrations profoundly influence outcomes such as enzyme yields.

Traditional methodologies such as Design of Experiments (DoE) and response surface methodology (RSM) dominate the field but necessitate extensive physical trials, imposing substantial costs and elongating timelines [1]. The advent of AI has revolutionized optimization by employing machine learning to streamline experimental designs with fewer trials [2]. Meanwhile, quantum-inspired computing enhances computational power and enables efficient search over large combinatorial spaces [3], and hybrid AI–quantum-inspired methodologies frequently outperform purely classical methods across diverse domains [4].

Quantum computing traces back to Ernst Ising's seminal work on spin models in 1925 [5]. Modern advances—including D-Wave One in 2011 and Fujitsu's DAU chip in 2018—demonstrate the potential of quantum and quantum-inspired hardware for solving QUBO problems at scale [6]. Classical Simulated Annealing (SA), a foundational technique for combinatorial optimization introduced by Kirkpatrick et al., remains one of the most widely used baseline methods for high-dimensional optimization [7]. Complementing this, Glover and Kochenberger introduced quantum-inspired optimization strategies that emulate quantum probability paths on classical hardware [8] and later formalized the QUBO framework as a universal modeling tool for NP-hard problems [9].

The practical relevance of quantum and quantum-inspired optimization has been demonstrated through benchmark studies in quantum annealing [10] and through quantum-enhanced applications such as factoring, scheduling, and constraint optimization [11].

In enzyme fermentation, different enzyme formulations contain varying Active Ingredient (AIN) values. The exponential growth in possible formulations, combined with the fact that each experiment requires over an hour, makes improving AIN levels through physical trials increasingly challenging. Software-driven approaches offer an alternative by reducing trial-and-error cycles, experiment counts, and development timelines.

This study applies a QUBO (Quadratic Unconstrained Binary Optimization) model to estimate the AIN in real-world fermentation processes. Due to the nonlinear nature of biological systems, quadratic models inevitably introduce distortions. Using the traditional Mean Square Error (MSE) cost function, experiments with 18 and 405 trials yielded errors of 19.99% and 15.51%, respectively—showing limited improvement despite increased data. In contrast, using the proposed Contour-Aware Cost Function reduced errors to 8.95% and 0.78% for the same experiments. The exceptional 0.78% error demonstrates that the QUBO model can predict

AIN with high precision for top-performing formulations, which is particularly meaningful for practical optimization.

## II. Related Work

Recent advances show the value of applying AI and machine learning to biological optimization, including enzyme production and bio-process modeling [12], and deep learning applications in genomics and biomedicine [13]. These advancements reinforce the importance of integrating computational models with biological experimentation.

ITRI (Industrial Technology Research Institute) focuses on three key domains in quantum technologies: quantum circuit simulation and analysis [14–15], quantum architecture and circuit design [16–17], and quantum applications [18] such as quantum neural networks (QNNs) and combinatorial optimization. This work [19] provides an integrated introduction to all three domains. Simulation and analysis are challenging because quantum states require $2^n$-scale matrix and vector operations, causing design complexity and analytical errors to grow exponentially. Circuit design also remains difficult, as leveraging entanglement and superposition offers parallelism but demands increasingly sophisticated architectures. On the application side, QNNs are widely explored for AI, while many industrial problems—such as risk evaluation, biomedical analytics, factory scheduling, warehouse management, materials optimization, network search, and AMR coordination—require strong combinatorial-optimization capabilities. These challenges underscore the importance of developing quantum technologies that can be effectively applied to real industrial scenarios and deliver meaningful quantum advantages.

To address these challenges, recent work in quantum-inspired optimization has demonstrated practical approaches that bridge theoretical capabilities with real-world applications. Recent progress in quantum-inspired optimization shows strong applicability across diverse domains. The digital QUBO solver introduced by Ou et al. [20] delivers high performance for large combinatorial problems on classical hardware. Similarly, the method in [21] applies quantum-inspired techniques and QAOA validation to achieve cost-effective fertilizer allocation in precision agriculture. In smart-mobility planning, the approach in [22] uses digital quantum annealing to determine optimal electric-vehicle charging-station placement with improved scalability. Meanwhile, the work in [23] extends quantum-inspired computing to transportation logistics by addressing the cruise passenger itinerary-scheduling problem, underscoring the versatility of these methods in complex real-world optimization tasks.

## III. Optimization process of enzyme fermentation

This study proposes a new method to optimize the enzyme fermentation process. The method consists of the following steps:

1. **Converting fermentation factors to binary**: First, the factors affecting fermentation (such as temperature, stirring frequency, pH value, tryptophan, brown rice powder, etc.) are converted into **22-bit** binary variables.
2. **Machine learning using known experimental data**: Based on existing experimental data, machine learning algorithms are used to optimize and generate QUBO (Quadratic Unconstrained Binary Optimization) coefficients. These coefficients include:
   - **Quadratic coefficients**: A total of 484 terms
   - **Linear coefficients**: A total of 22 terms
   - **Constant coefficient**: One term Thus, a total of 507 coefficients need to be optimized.
3. **Using QUBO software simulation**: QUBO software is used for simulation to find better fermentation formulas.
4. **Conducting more experiments**: Based on the simulation results, more targeted experiments are conducted. These experiments will repeat the above steps to further optimize the fermentation formula.

### 3.1 Expectation Results of Simulated Annealing (SA)

Assuming there are $2^{100K}$ possible states, the cost function is approximately Gaussian distributed, as shown in Fig 1, where the x-axis represents the normalized cost and the y-axis represents the frequency of occurrence. The following are the simulation results and conclusions for different scenarios:

*Scenario 1: Manufacturer A's Guess*
1. Assume Manufacturer A can guess a state with a cost of $-\sigma$.
2. Using SA, if a randomly guessed state has a cost lower than $-\sigma$, it is considered a success; otherwise, it is a failure.
3. According to the Gaussian distribution, the failure rate for the first iteration is $68\% + \frac{1-68\%}{2} = 84\%$.
4. To reduce the failure rate to less than 0.01 after n iterations, it must satisfy $0.84^n < 0.01$. Calculating this gives $n > 26.4$, meaning that running 27 iterations of SA has a 99% chance of achieving a result better than Manufacturer A.

*Scenario 2: Manufacturer B's Guess*
1. Assume Manufacturer B uses a heuristic to find a solution with a cost of $-2\sigma$.
2. Similarly, using SA, the cost of the randomly guessed state must be lower than $-2\sigma$ to be considered a success.
3. The failure rate for the first iteration is $95.44\% + \frac{1-95.44\%}{2} = 97.72\%$.
4. To reduce the failure rate to less than 0.01 after n iterations, it must satisfy $0.9772^n < 0.01$. Calculating this gives $n > 199.67$, meaning that running 200 iterations of SA has a 99% chance of achieving a result better than Manufacturer B.

The effectiveness of SA is related to the data distribution (proximity of the minimum value) and the ability of heuristic algorithms. When seeking a solution that is better than the original, the time required for simulation increases exponentially (by a factor of $e^2 = 7.398$) for each improvement of σ.

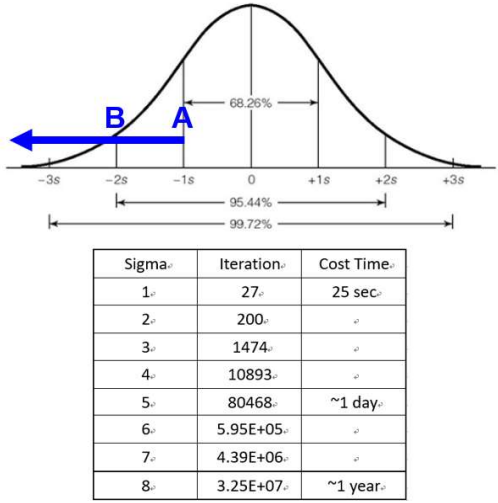

**Fig. 1.** The cost function Gaussian distribution.

Under a Gaussian assumption of the cost distribution, for problem sizes of $n=100$ and $n=1000$ bits, the number of samples required to obtain solutions of comparable quality is similar. Consequently, the required computation time does not grow exponentially with respect to $n$. We define the optimization depth $m$ as $m = -(c-u)/\sigma$, where $c$ denotes the optimal cost, $u$ is the mean of the cost distribution, and $\sigma$ is the standard deviation. Since the optimal cost $c$ is typically lower than the mean $u$, the value of $m$ is positive. In SA, the thermal parameter is chosen to target a specific optimization depth $m$. Each exploration step requires a QUBO cost evaluation with time complexity $O(n^2)$. As a result, the overall time complexity of SA can be approximated as $O(n^2 \times e^{2m})$.

When the thermal parameter is fixed, the optimization depth $m$ becomes bounded and can therefore be treated as a constant and omitted in Big-O notation. For QUBO problems, compared with brute-force search of complexity $O(2^n)$, SA reduces the computational complexity to approximately $O(n^2)$. The Fujitsu Digital Annealer (DAU) further reduces this complexity to $O(n)$. Nevertheless, achieving higher-quality optimization results still requires exponentially increasing runtime. For example, extending the computation time from 1 day to 7 days may improve the solution quality by only about one standard deviation.

### 3.2 Walking in the Snow in Search of Plum Blossoms Algorithm

An improved data augmentation and elimination method is proposed to enhance the active ingredients (AIN) in enzyme fermentation. As shown in Fig. 2, the specific methods are as follows:

● Data Augmentation

Due to the large number of QUBO coefficients ($O(n^2)$) and the limited amount of experimental data (only 18 samples), overfitting may occur, leading to anomalous AIN values. To mitigate this issue, we increased the number of training samples by adding $n^2$ augmented formulations. The AIN values of these augmented samples are randomly assigned slightly below the average AIN, which stabilizes the AIN distribution during model training.

● Data Elimination

After generating new experiments, the augmented samples whose Hamming distance from neighboring points is less than or equal to 3 are automatically discarded to prevent them from overshadowing potentially better formulations during the data augmentation process. In this process, the original experimental data are preserved, ensuring the reliability and completeness of the core dataset.

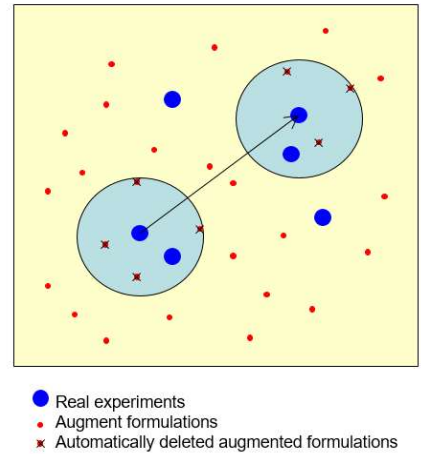

**Fig 2.** Walking in the snow in search of plum blossoms

The left part of Fig. 3 illustrates the standard AI and quantum-inspired optimization workflow. The middle part represents a scenario in which the QUBO model fails to recommend new formulations because all candidate solutions have already been experimentally evaluated. In this case, additional experiments are conducted on neighboring formulations, defined as those with a Hamming distance of one from the current best formulation. The optimization process then returns to the standard workflow shown on the left. This iterative procedure continues until none of the neighboring formulations yields further improvement in AIN and the QUBO model can no longer provide superior recommendations. At this point, the framework is terminated, as the system is considered to have identified the optimal formulation.

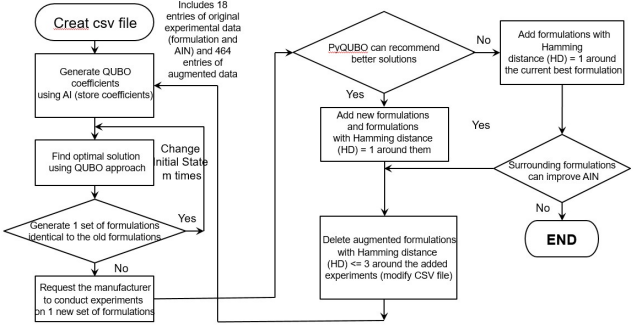

**Fig 3.** Proposed Simulation Flowchart

### 3.3 Improvement of Machine Learning

The following section investigates how to improve the accuracy of the QUBO model under ML-based coefficient optimization. To expedite the discovery of the formula that maximizes enzyme fermentation AIN, different ML strategies are employed to minimize the error between the real AIN and the estimated AIN of the best-performing formula. Based on experimental observations in Fig 4, the choice of the initial guess plays a critical role in minimizing the QUBO estimation error.

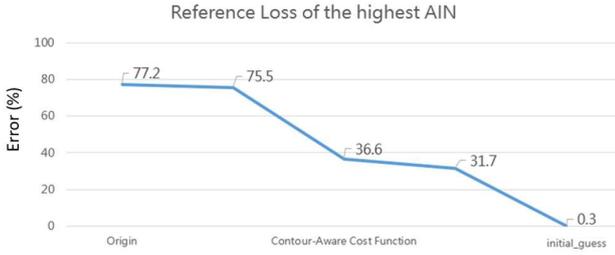

**Fig 4.** Improvement of Machine Learning

- Coarse-Fine Tuning Method

A two-stage coarse-fine tuning strategy is proposed for QUBO coefficient optimization to improve convergence. In the first stage, only the linear and constant coefficients of the QUBO model are optimized using the training dataset. In the second stage, all coefficients are jointly optimized with the results from the first stage serving as the initial values. Experimental results demonstrate that the proposed coarse-fine tuning strategy significantly reduces the error of the QUBO model.

- Contour-Aware Cost Function

The relationship between enzymatic fermentation formulation parameters and the resulting AIN is inherently nonlinear. When a QUBO model is used to approximate such nonlinear input–output behavior, substantial modeling errors are generally unavoidable. As a result, it is commonly believed that a QUBO-based approach is unlikely to identify improved formulations in this setting. In contrast to this prevailing assumption, we propose an alternative perspective. Rather than requiring the QUBO model to achieve uniform accuracy across all formulations, we suggest that the model should exhibit higher accuracy for high-quality formulations and lower accuracy for inferior ones. Such a deliberately biased accuracy distribution is not only sufficient but also advantageous for steering the search process toward improved formulations.

Based on this insight, we propose a contour-aware cost function defined as

$$\text{Contour Aware Cost Function} = \frac{(E-H)^2}{e^{\frac{Max-H}{100}}} \ldots Eq(1)$$

where $Max$ denotes the maximum AIN value observed across all experiments, $H$ represents the AIN value obtained from a specific experiment, and $E$ denotes the AIN estimated by the QUBO model.

By interpreting $H$ across different experimental outcomes as contour levels, the proposed cost function assigns varying importance to modeling accuracy. When ($Max-H$) is small, indicating a high-quality formulation, the exponential term introduces minimal attenuation, and modeling accuracy is strongly emphasized. In contrast, when ($Max-H$) is large, corresponding to a low-quality formulation, the cost decays exponentially, reflecting the reduced importance of accurate modeling in these regions. This contour-aware weighting mechanism allows the QUBO model to concentrate its representational capacity on formulations with higher AIN values, thereby facilitating the identification of improved formulations.

## IV. EXPERIMENTAL RESULTS

In optimizing enzyme fermentation formulations (x0–x21) and their corresponding active ingredient (AIN), previous blind experimentation exceeding 600 trials achieved a maximum AIN of only 8,481. Although adjusting the formulations provides opportunities to increase the AIN, the vast number of possible combinations makes identifying improved formulations challenging. Starting from 18 initial experiments, we conducted incremental experiments as follows: (1) generating a QUBO model through ML-based optimization using the available experimental data, (2) using QUBO simulation to suggest the next candidate formulation, and (3) performing an enzyme fermentation experiment based on the suggested formulation. Using this approach, our method reached an AIN of 10,068 with only 405 experiments, demonstrating its efficiency in guiding formulation optimization.

Table 1 summarizes the experimental results, with the iteration count incremented only when an experiment yielded an increased AIN. For instance, Iteration = 2 at Experimental Number = 41 indicates that the 41st experiment improved the AIN, advancing the iteration from 1 to 2, while Iteration = 3 at Experimental Number = 64 reflects a further improvement, incrementing the iteration from 2 to 3. This approach highlights the experiments that directly contributed to performance gains. Figure 5 illustrates the corresponding AIN improvements based on Table 1.

Figure 6 compares the prediction errors obtained with and without the contour-aware cost function. Mean square error (MSE) represents the error calculated using the commonly used cost function, showing that without the contour-aware cost function, the error remains relatively high (about 14%~20%). In contrast, contour-aware error (CAE) represents the error based on the contour-aware cost function, which is significantly reduced from 8.95% to 0.78%. The substantially lower error implies that the QUBO model can accurately predict the AIN for high-performing formulations and is capable of suggesting the next candidate formulations to maximize AIN.

Table 1. Experiments based on the proposed method

| Iteration | Number of Experiments | The recommended new recipe (Best_solution) | Real AIN | Extimate AIN | MSE(%) | Contour-Aware MSE(%) |
|---|---|---|---|---|---|---|
| 1 | 18 | [0, 1, 0, 1, 1, 1, 1, 0, 1, 1, 1, 1, 1, 1, 1, 1, 1, 1] | 8481 | 6584 | 19.99 | 8.95 |
| 2 | 41 | [0, 0, 0, 0, 0, 1, 1, 0, 0, 1, 1, 1, 0, 1, 1, 1, 1, 1] | 8867 | 8775 | 17.63 | 8.26 |
| 3 | 64 | [0, 0, 0, 0, 0, 1, 1, 0, 0, 1, 1, 1, 0, 1, 1, 0, 1, 1] | 8875 | 9080 | 16.81 | 7.67 |
| 4 | 87 | [0, 0, 0, 0, 0, 1, 1, 0, 0, 1, 1, 1, 1, 1, 1, 0, 1, 1] | 8881 | 9164 | 16.19 | 7.17 |
| 5 | 110 | [0, 0, 0, 0, 0, 1, 1, 0, 0, 1, 1, 0, 1, 0, 1, 0, 1, 1] | 9068 | 9185 | 15.74 | 6.68 |
| 6 | 133 | [0, 0, 0, 0, 0, 1, 1, 0, 0, 1, 1, 0, 1, 0, 1, 1, 1, 1] | 9049 | 9316 | 15.09 | 5.96 |
| 7 | 156 | [0, 0, 0, 0, 0, 1, 0, 0, 1, 1, 0, 1, 1, 0, 1, 0, 1, 1] | 9248 | 9333 | 14.41 | 5.35 |
| 8 | 179 | [0, 0, 0, 0, 0, 1, 1, 0, 0, 1, 1, 0, 1, 1, 0, 1, 0, 1] | 9275 | 9441 | 13.97 | 4.77 |
| 9 | 202 | [0, 0, 0, 0, 0, 1, 0, 0, 1, 1, 0, 1, 0, 0, 1, 0, 1, 1] | 9382 | 9547 | 13.67 | 4.20 |
| 10 | 224 | [0, 0, 0, 0, 0, 0, 1, 0, 0, 1, 1, 0, 1, 0, 0, 1, 0, 1] | 9466 | 9695 | 13.59 | 3.58 |
| 11 | 247 | [0, 0, 0, 0, 0, 0, 1, 0, 0, 1, 1, 0, 0, 0, 0, 0, 1, 0] | 9480 | 9738 | 13.55 | 2.91 |
| 12 | 270 | [0, 0, 0, 0, 0, 0, 1, 1, 1, 0, 1, 0, 1, 0, 0, 1, 0, 1] | 9454 | 9861 | 13.82 | 2.18 |
| 13 | 293 | [0, 0, 0, 0, 0, 0, 0, 1, 1, 0, 1, 0, 1, 0, 1, 0, 1, 0] | 9461 | 9877 | 14.01 | 1.52 |
| 14 | 315 | [0, 0, 0, 0, 0, 0, 1, 0, 0, 1, 0, 1, 0, 1, 1, 1, 0, 0] | 9767 | 10027 | 14.05 | 1.19 |
| 15 | 338 | [0, 0, 0, 0, 0, 0, 0, 0, 1, 0, 0, 0, 1, 0, 1, 1, 0, 1] | 9748 | 10172 | 14.73 | 0.82 |
| 16 | 361 | [0, 0, 0, 0, 0, 0, 0, 1, 0, 0, 0, 1, 1, 0, 1, 0, 0, 1] | 9756 | 10156 | 15.05 | 0.79 |
| 17 | 384 | [0, 0, 0, 0, 0, 0, 0, 1, 0, 0, 0, 0, 1, 0, 1, 1, 0, 0, 1] | 9957 | 10148 | 15.32 | 0.78 |
| 18 | 405 | [0, 0, 0, 0, 0, 0, 0, 0, 0, 0, 0, 0, 0, 1, 0, 1, 1, 0, 0, 0] | 10068 | 10171 | 15.51 | 0.78 |

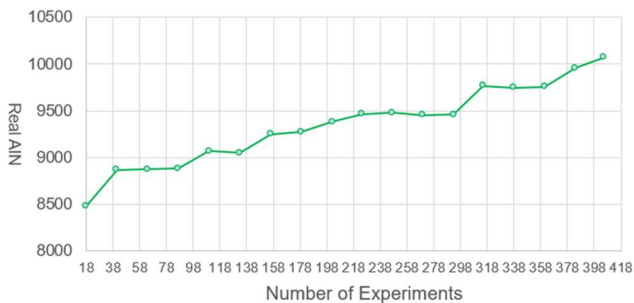

**Fig 5.** Real Active Ingredients (AIN)

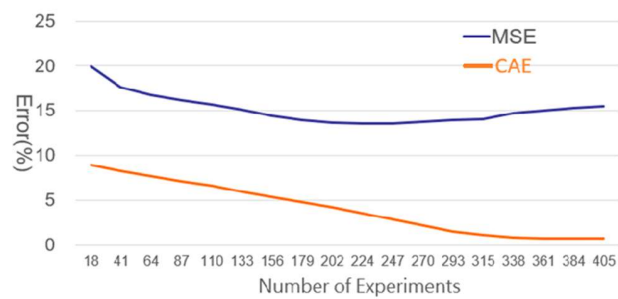

**Fig 6.** QUBO model prediction error for AIN

## V. CONCLUSION

This paper presents a novel framework that integrates AI with quantum-inspired techniques for solving combinatorial optimization problems. Experimental results on enzyme fermentation demonstrate the effectiveness of the proposed ML-based QUBO framework in identifying the better formulation for enhancing active ingredients. In this framework, quantum-inspired techniques play an auxiliary role by accelerating the QUBO process. More importantly, the framework provides a systematic pathway for reconstructing QUBO models, making it broadly applicable to the domains with limited prior knowledge.

The proposed framework offers several key advantages: (1) as the number of experiments increases, the AI model improves in accuracy, enabling the QUBO model to recommend better formulations and thereby establishing a positive feedback loop; (2) the iterative integration of AI and quantum-inspired techniques alleviates the difficulty of QUBO model construction which traditionally requires extensive domain expertise; and (3) considerable time and cost savings are achieved by replacing some physical experiments with software simulations.

Employing the proposed method, the actual AIN increased from 8,481 to 10,068, an improvement of 1,587, while the number of experiments was reduced from over 600 to 405. With the expanded dataset, the MSE decreased from 19.99% to 15.51%, and the contour-aware error metric was reduced from 8.95% to 0.78%. This case study highlights both the complexity of the optimization process and the substantial performance gains achieved.

In summary, this approach boosts the efficiency of optimizing enzyme fermentation formulations and showcases the potential of AI and quantum-inspired techniques in solving complex problems, achieving high-quality results faster and more cost-effectively.

REFERENCES

[1] D. C. Montgomery, Design and Analysis of Experiments, 9th ed. Hoboken, NJ, USA: Wiley, 2017.
[2] Y. Bengio, I. Goodfellow, and A. Courville, Deep Learning. Cambridge, MA, USA: MIT Press, 2016.
[3] T. Albash and D. A. Lidar, "Adiabatic quantum computation," Rev. Mod. Phys., vol. 90, no. 1, p. 015002, 2018.
[4] A. Lucas, "Ising formulations of many NP problems," Front. Phys., vol. 2, p. 5, 2014.
[5] E. Ising, "Contributions to the theory of ferromagnetism," Z. Phys., vol. 31, no. 1, pp. 253–258, 1925.
[6] M. W. Johnson et al., "Quantum annealing with manufactured spins," Nature, vol. 473, no. 7346, pp. 194–198, 2011.
[7] S. Kirkpatrick, C. D. Gelatt Jr., and M. P. Vecchi, "Optimization by simulated annealing," Science, vol. 220, no. 4598, pp. 671–680, 1983.
[8] F. Glover and G. Kochenberger, "Quantum-inspired probability paths and large-scale optimization," Comput. Inf. Syst., vol. 10, no. 1, pp. 1–12, 2006.
[9] F. Glover, G. Kochenberger, and Y. Du, "A tutorial on formulating and using QUBO models," arXiv preprint arXiv:1811.11538, 2018.
[10] C. C. McGeoch, "Optimization benchmarks for quantum annealing," in Proc. ACM Int. Conf. High Performance Computing, 2013.
[11] D. Venturelli, S. Mandrà, S. Knysh, et al., "Quantum annealing for prime factorization," Phys. Rev. X, vol. 5, no. 3, Art. no. 031040, 2015.
[12] R. S. Singh et al., "Recent advances in enzyme production and fermentation optimization using machine learning," Biotechnol. Adv., vol. 37, no. 8, Art. no. 107431, 2019.
[13] J. Zou, M. Huss, et al., "A primer on deep learning in genomics and biomedicine," Nat. Biotechnol., vol. 37, pp. 1091–1097, 2019.
[14] Y.-T. Kao, H.-R. Lu, Y.-J. Chang, and D.-W. Lu, "Quantum-Chiplet: A novel Python-based efficient and scalable design methodology for quantum circuit verification and implementation," arXiv preprint arXiv:2503.10054, 2025.
[15] H.-R. Lu, Y.-T. Kao, Y.-J. Chang, J. Young, and D.-W. Lu, "Quantum-Chiplet: A VLSI-like methodology for hierarchical quantum design and verification," in Proc. IEEE Int. Symp. Very Large Scale Integration (ISVLSI), Kalamata, Greece, Jul. 2025.

[16] Y.-T. Kao, Y.-J. Chang, and Y.-W. Tseng, "Mixed-signal quantum circuit design for option pricing using Design Compiler," arXiv preprint arXiv:2506.15936, 2025.

[17] Y.-T. Kao, Y.-J. Chang, and Y.-W. Tseng, "Novel quantum circuit designs of random injection and payoff computation for financial risk assessment," arXiv preprint arXiv:2507.23310, 2025.

[18] C.-Y. Liu et al., "Training classical neural networks by quantum machine learning," in Proc. IEEE Int. Conf. Quantum Comput. Eng. (QCE), pp. 34–38, 2024.

[19] Y.-W. Tseng et al., "Achieving sub-exponential speedup in gate-based quantum computing for QUBO," arXiv preprint arXiv:2510.15334, 2025.

[20] C. H. Ou, P.-Y. Chu, T. H. Tsai, Y. Chen, and C.-Y. Chen, "A high-performance digital QUBO solver for quantum-inspired optimization," IEEE Nanotechnol. Mag., accepted for publication, Nov. 2025.

[21] C. H. Ou, Y.-W. Kan, M. Ohzeki, and C.-Y. Chen, "Quantum-inspired optimization and QAOA validation for cost-effective fertilizer allocation in precision agriculture," in Proc. IEEE Int. Conf. Quantum Comput. Eng. (QCE 2025), Sept. 2025.

[22] C. H. Ou, C.-C. Cheng, C.-Y. Chen, K. Bhumkittipich, S. Romphochai, "Solving Optimal Electric Vehicle Charging Station Placement Problem Using Digital Quantum Annealing," IEEE Journal of Communications and Networks, vol. 27, no. 4, pp. 252-263, Aug. 2025.

[23] C. H. Ou, Y.-C. Lin, W.-Y. Chen, and W.-C. Wu, "Solving the cruise passenger itinerary scheduling problem using quantum-inspired computing," in Proc. IEEE Int. Conf. Quantum Commun., Netw., Comput. (QCNC), Mar. 2025.